\documentclass[letterpaper, 10 pt, conference]{ieeeconf}  % Comment this line out
\IEEEoverridecommandlockouts                              % This command is only
\usepackage{balance}
\usepackage{cite}
\usepackage{cite}
\usepackage{svg}
\usepackage{amsthm}
\usepackage{amssymb}

\newtheorem{Problem}{Problem}[]
\newtheorem{Definition}{Definition}[]
\newtheorem{Assumption}{Assumption}[]
\newtheorem{Lemma}{Lemma}[]
\usepackage{xcolor,soul,framed} %,caption
\usepackage{graphicx}
\usepackage{todonotes}
\usepackage[cmex10]{amsmath}
\usepackage{array}
\usepackage{algorithm}
\usepackage{algpseudocode}
\usepackage{amsfonts}
\usepackage{tabularx} %for automation of column widths so that the table spans the width defined by textwidth and linewidth
\usepackage{amsfonts} 
\usepackage{hyperref} 
\usepackage{tcolorbox}
\usepackage{algorithmicx}
\algnewcommand\algorithmicforeach{\textbf{for each}}
\algdef{S}[FOR]{ForEach}[1]{\algorithmicforeach\ #1\ \algorithmicdo}

\DeclareMathOperator*{\argmin}{arg\,min}
\title{\LARGE \bf
A Modular Safety Filter for Safety-Certified Cyber-Physical Systems
}
\author{Mohammad Bajelani, Mehran Attar, Walter Lucia and Klaske van Heusden
\thanks{We acknowledge the support of the Natural Sciences and Engineering Research Council of Canada (NSERC) [RGPIN-2023-03660].}% <-this % stops a space
\thanks{Mohammad Bajelani and Klaske van Heusden are with the University of British Columbia, School of Engineering, 3333 University Way, Kelowna, BC V1V 1V7, Canada. Mehran Attar and Walter Lucia are with the Concordia Institute for Information Systems Engineering (CIISE), Concordia University, Montreal, QC, H3G 1M8, Canada
{\tt\small mohammad.bajelani, klaske.vanheusden @ubc.ca, mehran.attar, walter.lucia @concordia.ca}}}
\begin{document}
\maketitle
%%%%%%%%%%%%%%%%%%%%%%%%%%%%%%%%%%%%%%%%%%%%%%%%%%%%%%%%%%%%%%%%%%%%%%%%%%%%%%%%
\begin{abstract}

Nowadays, many control systems are networked and embed communication and computation capabilities. Such control architectures are prone to cyber attacks on the cyberinfrastructure. Consequently, there is an impellent need to develop solutions to preserve the plant's safety against potential attacks. To ensure safety, this paper introduces a modular safety filter approach that is effective for various cyber-attack types. This solution can be implemented in combination with existing control and detection algorithms, effectively separating safety from performance. The safety filter does not require information on the received command's reliability or the anomaly detector's feature. It can be implemented in conjunction with high-performance, resilient controllers to achieve both high performance during normal operation and safety during an attack. As an illustrative example, we have shown the effectiveness of the proposed design considering a multi-agent formation task involving 20 mobile robots. The simulation results testify that the safety filter operates effectively during undetectable, intelligent attacks.

\end{abstract}

%%%%%%%%%%%%%%%%%%%%%%%%%%%%%%%%%%%%%%%%%%%%%%%%%%%%%%%%%%%%%%%%%%%%%%%%%%%%%%%%
\section{INTRODUCTION}

Cyber-Physical Systems (CPS) are networked control systems with a tight integration with computation and communication capabilities\cite{pasqualetti2015control}. Merging cyber technologies with physical systems significantly boosts operational efficiencies. However, it also introduces vulnerabilities that undermine the reliability of essential infrastructure as the communication lines present opportunities for hackers to manipulate data lines and initiate cyber attacks. Various solutions have been proposed to prevent, detect, and mitigate cyber-attacks using control theoretical tools\cite{dibaji2019systems}. Most of the solutions consider systems without constraints, and/or the mitigation strategies rely on the use of an anomaly detector to ensure that the system does not enter unsafe configurations.

Recently, there has been an increasing trend in constrained CPS to address safety concerns explicitly. This involves formally defining safe zones as constraints within state and input spaces. In this paper, a Modular Safety Filter (MSF), inspired by safety-certified learning-based controllers \cite{wabersich2023data}-\cite{hsu2023safety}, is proposed to satisfy safety constraints in the presence of cyber-attacks on the actuator and sensor signals. Due to its modularity, this method can be used as a standalone technology alongside other resilient controllers and anomaly detectors. Performance and safety criteria are separated in our architecture, simplifying the design process. Since MSF makes no assumptions about the attacked signal, such as limited bandwidth and small bounds or the attacker's computational power, it can be used in a wide range of situations.

Prior work largely focuses on linear systems. In \cite{franze2023cyber} and \cite{yang2023resilient}, Model Predictive Control (MPC) is employed for linear systems under False Data Injection (FDI) attacks, guaranteeing the stability and constraint satisfaction of the system. In \cite{wei2024resilient}, a distributed MPC and attack detection framework is proposed for constrained linear multi-agent systems under adversarial attacks. In \cite{gheitasi2022worst}, a tracking method that requires reachable sets is proposed for constrained linear systems under arbitrary attacks on both the actuation and measurement lines. A data-driven approach for LTI systems is proposed in \cite{attar2024data2} and \cite{liu2022data}, introducing a safety verification plus emergency control module, assuming only noise-polluted input-state trajectories are available. A semi-definite approach, assuming bounded additive attacker's signals, is proposed in \cite{escudero2023safety} to design a safety-preserving filter for deterministic LTI systems under FDI attacks. In \cite{lucia2022supervisor}, a set-theoretic receding horizon control has been proposed to address FDI and denial of service attacks for LTI systems. In \cite{gheitasi2021safety} and \cite{zhang2020reachability}, reachability analysis is used and investigated to design safety-preserving platforms for LTI systems. A solution for nonlinear systems is proposed in \cite{lin2023secondary} and \cite{al2023timing}, which provides safety based on an invariant set of SOS-based Lyapunov functions, resulting in conservative ellipsoidal safe sets.

The majority of proposed methods rely on reachable set arguments, which often restrict their focus to LTI systems, as computing reachable sets becomes challenging or costly for high-order and nonlinear systems. Additionally, many methods aim to address both control performance and safety, requiring a balance between these objectives and computational cost — a challenge for MPC, particularly with long horizons in high-order systems. In contrast, a modular safety filter approach ensures safety without modifying the existing system or requiring long prediction horizons, making it computationally efficient and practical for nonlinear systems. This paper explores how CPS safety can be maintained without altering existing components, such as controllers and anomaly detectors, by incorporating a minimally invasive filter. MSF facilitates the integration of established control methods, achieving both performance and safety without system modifications.

\begin{figure}[t] 
    \centering
    \includegraphics[width=0.65\linewidth, trim=0cm 2.5cm 0cm 1.5cm, clip]{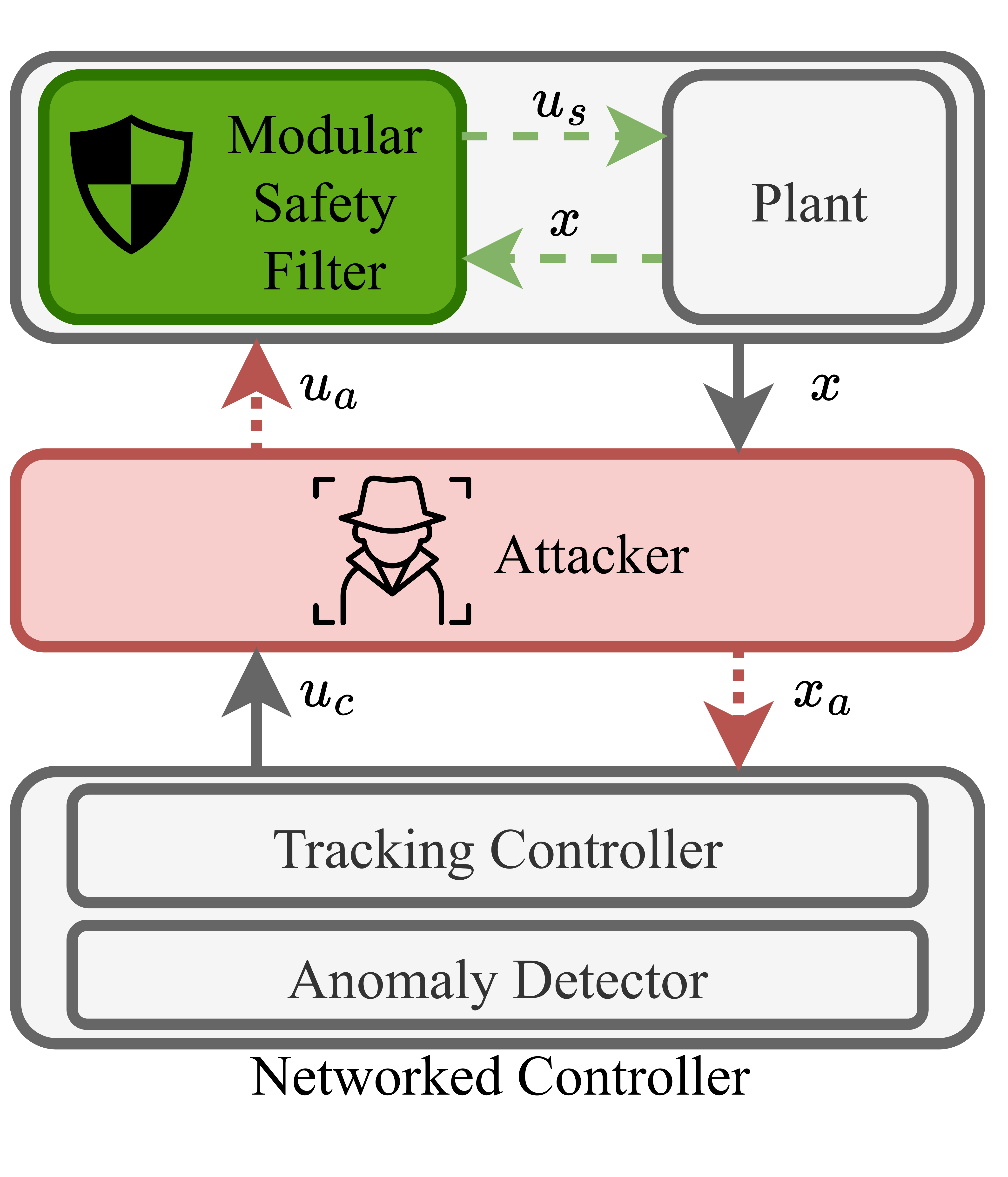}
    \caption{Proposed safety-certified architecture for cyber-physical systems. The green dashed lines (\protect\includegraphics[width=2.5em, height=1em]{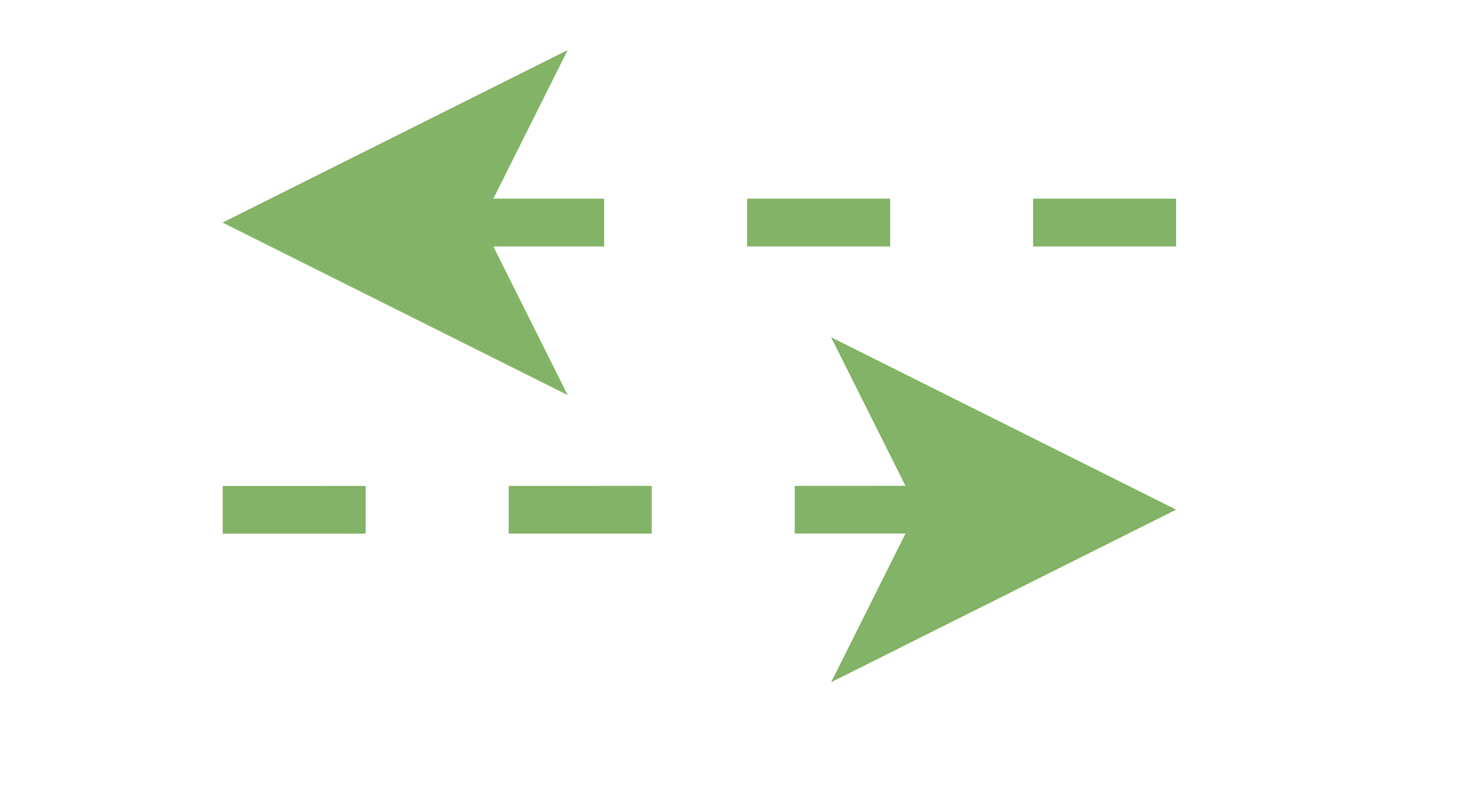}) represent the communication between the modular safety filter and the plant, assumed to be unaffected by network attacks. The attacker can target communication channels between the plant and the networked controller, represented by gray solid lines (\protect\includegraphics[width=2.5em, height=1em]{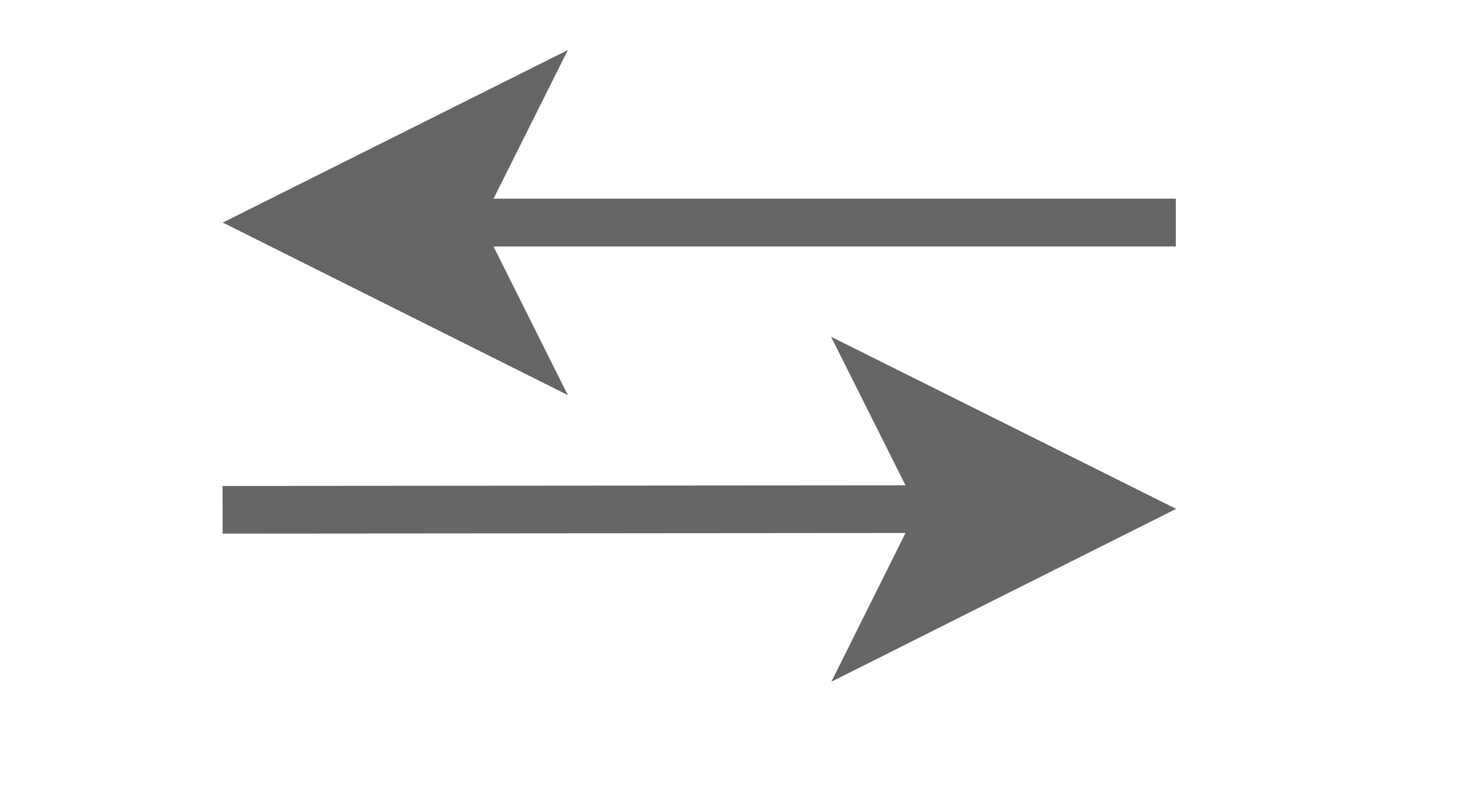}), and injects malicious signals through red dotted lines (\protect\includegraphics[width=2.5em, height=1em]{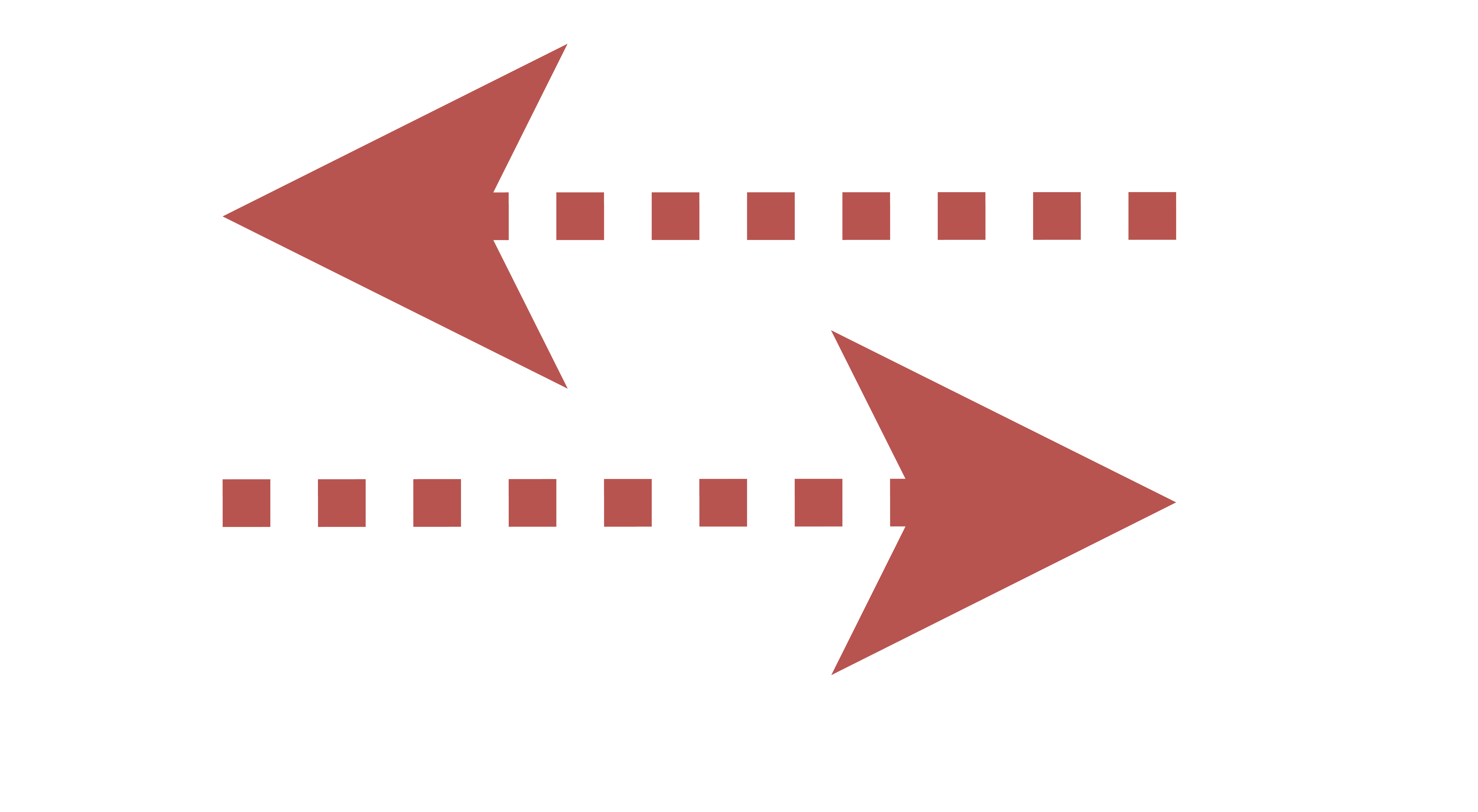}). As an illustrative example, this architecture can be applied to a robotic scenario, where the networked controller communicates with the robot via Wi-Fi, with the onboard modular safety filter operating, leaving the Wi-Fi connection vulnerable to potential threats. See the following color box for a detailed discussion of the proposed architecture.}
    \label{fig: Proposed_Architecture}
\end{figure}
\begin{tcolorbox}[float,floatplacement=!t, colframe=black!50!green, colback=white, coltitle=white, title=\textbf{Why safety filters must be local to the plant.}]
It is important to emphasize that a covert, intelligent attack on communication channels may evade detection by any anomaly detectors on the networked controller's side \cite{smith2015covert}. This underscores the need for a safety filter to act as a \emph{localized policy} on the plant side. In CPS domains like the smart grid, the networked controller oversees and synchronizes subsystems to achieve a unified goal while managing only each subsystem's reference/input signal; see discussions in \cite{attar2024data} and \cite{escudero2023safety}. Therefore, the safety filter requires access to the unaltered states, which can only be achieved through a localized implementation. However, in cases where the local network faces a sophisticated, intelligent attack, modifications to the filter may be necessary to maintain safety \cite{arnstrom2024stealthy}. Finally, suppose an attacker executes a coordinated, intelligent attack. In that case, the only effective defense for the system is to implement an emergency safety module, such as the modular safety filter, \emph{local to the plant}.
\end{tcolorbox}

The remainder of the paper is organized as follows. Preliminary material, adversarial capabilities, and the problem statement are described in section \ref{Preliminary Material and Problem Statement}. Section \ref{Section: Modular Safety Filter} includes the proposed MSF. A multi-agent setup consisting of 20 mobile robots is described in section \ref{Modular Predictive Safety Filter for Multi-Agent Mobile Robot System}. Simulation results considering undetectable, intelligent attacks are presented in Section \ref{section: Numerical Results}. Lastly, conclusions and limitations of the present work are discussed in section \ref{Discussion and Concluding Remarks}.

\section{Preliminary Material and Problem Statement} \label{Preliminary Material and Problem Statement}

This section provides an overview of the CPS problem considered in this paper, the proposed MSF architecture, and the networked controller depicted in Fig. \ref{fig: Proposed_Architecture}. It also details the formulation of the plant's dynamics and the types of cyber-attacks used for simulation examples.

\subsection{Plant's Dynamics}
Consider a class of discrete-time dynamical systems that can be described by a set of nonlinear equations as follows,
\begin{subequations} \label{equ: nonlinear}
\begin{equation} \label{equ: nonlinear_dynamics}
    x(t+1) = f(x(t), u(t)),  \end{equation}
\begin{equation}\label{equ: constraints} 
    u(t) \in \mathcal{U}, \; x(t) \in \mathcal{X}, 
    \end{equation}
\end{subequations}
where \( t \in \mathbb{N} \) is the time step, \( x(t) \in \mathbb{R}^n \) the state, \( u(t) \in \mathbb{R}^m \) the control input, \( \mathcal{U} \) the input constraints, and \( \mathcal{X} \) the state constraints. The function \(f(x(t), u(t))\) is a generic function describing the plant’s dynamic.
\begin{Definition}[Safety]\label{Def: Safety}  
The dynamical system \eqref{equ: nonlinear} is said to be safe if the input-state pair \((u(t), x(t))\) satisfies \((u(t), x(t)) \in \mathcal{U} \times \mathcal{X}\) for all \(t \geq 0\).  
\end{Definition}

\begin{Definition}[Safe Control Invariant Set] \label{Def: Safe Set}
    A safe control invariant set, $\mathcal{S} \subseteq \mathcal{X}$, is a set of initial states $x(t)$ such that there exists a control input $u(t) \in \mathcal{U}$ ensuring that $x(t+1) \in \mathcal{S}$ for all $t>0$. Formally:
    \begin{equation} \label{equ: safe set def}
        \mathcal{S} = \{x(t) \in \mathcal{X} \mid \exists u(t) \in \mathcal{U}, \ x(t+1) \in \mathcal{S}, \forall t>0\}.
    \end{equation}
\end{Definition}

\subsection{Adversarial Capabilities}\label{Attacker Abilities}

Let us assume that $x(t)$ is sent from a local network to the networked controller via an unsecured network. The networked controller computes the control action, $u_c(t)$, and transmits it back to the local network. These signals are susceptible to cyber-attacks, denoted by $x_a(t)$ and $u_a(t)$, respectively. Without losing generality, a cyber attack can be described using the following unknown function:
\begin{equation} \label{eq: attacker h}
        (u_a(t),x_a(t)) = h(u_c(t),x(t)).
\end{equation}
Function $h(u_c(t),x(t))$ can be defined to represent different types of attack and adversarial capabilities. Since the proposed modular safety filter does not rely on assumptions on the attack, i.e., the function $h(u_c(t),x(t))$ is unknown to the proposed safety filter. We will consider two situations in the illustrative simulation example in Section \ref{section: Numerical Results}: attack-free and intelligent attacks.

\subsubsection{Attack-Free Scenario}  
In the attack-free scenario, \( h(u_c(t), x(t)) \) is the identity function, indicating that the attacker cannot alter the signals \( x(t) \) and \( u_c(t) \):
\begin{equation} \label{equ: attack-free}
    u_a(t) = u_c(t), \quad x_a(t) = x(t),
\end{equation}
In other words, the networked controller receives unaltered \( x(t) \) from the local network, and the local network receives unaltered \( u_c(t) \) from the networked controller.

\subsubsection{Intelligent Attack}
The attacker is assumed to know the system dynamics, disclosure, and disruptive resources on the data transmitted, $x(t)$ and $u_c(t)$; undetectable covert attacks can be launched. Furthermore, the attacker has sufficient computational power to compute $h(u_c(t),x(t))$ resorting to any desired optimal policy. For this attack, it has been proved that no anomaly detector - whether implemented as an active or passive module - located on the networked controller’s side can detect its presence \cite{7011176}. In particular, for an FDI attack, we assume that the attacker can introduce arbitrary perturbations, \(\delta_x(t)\) and \(\delta_u(t)\), to the control input and state measurement vectors:
\begin{equation} \label{equ: FDI}
    u_a(t) = u_c(t) + \delta_u(t), \quad x_a(t) = x(t) + \delta_x(t).
\end{equation}

\subsection{Networked Controller} \label{Networked Controller}
A networked controller typically consists of two main components: a tracking controller and an anomaly detector. The tracking control policy can also be formulated to address tasks such as regulation, tracking, or other objectives. It is generally expressed as:
\begin{equation} \label{tracking controller}
        u_c(t) = g(r(t), x_a(t)),
\end{equation}
where \( r(t) \) represents the reference trajectory. The tracking controller is assumed to be safety-certified under the attack-free condition in \eqref{equ: attack-free}. Note that violating this assumption does not compromise the plant's safety, but it may trigger false alarms in anomaly detection systems (see Remark 2). Therefore, the function \( g(r(t), x_a(t)) \) is unknown to the proposed safety filter\footnote{The attacker may exploit previously recorded inputs and states, as seen in replay buffer attacks, while the tracking controller may use future reference and past input-state data. For simplicity, this notation is omitted here.}.
We consider a passive binary anomaly detection mechanism designed to detect cyber-attacks. The anomaly detector is described as follows:
\begin{equation} \label{eq: anomaly_general}
    a(t) = \mathcal{A}(\{x_a(i)\}_{i=t_0}^{t}, \{u_c(i)\}_{i=t_0}^{t}),
\end{equation}
where \( a(t) \in \{0, 1\} \), with \( a = 0 \) and \( a = 1 \) indicating the absence and presence of an anomaly, respectively. Here, \( \mathcal{A}(\{x_a(i)\}_{i=t_0}^{t}, \{u_c(i)\}_{i=t_0}^{t}) \) is a generic function that can store an arbitrarily large history of the signals \( x_a(i) \) and \( u_c(i) \), where \( 0 \leq t_0 \leq t \). It is important to note that in the case of an intelligent attack, no anomaly detection mechanism can reliably detect the anomaly on the networked controller side. Therefore, the anomaly detector in \eqref{eq: anomaly_general} is only included for completeness and is intended for use in simulation examples.

\begin{Problem}\label{Problem}
    Design a local control policy to ensure that system \eqref{equ: nonlinear} remains safe, as defined in Definition \ref{Def: Safety}, regardless of the attacker strategy \eqref{eq: attacker h}. The policy should also preserve the functionality of the tracking controller \eqref{tracking controller} and anomaly detector \eqref{eq: anomaly_general} under attack-free conditions \eqref{equ: attack-free}.
\end{Problem}

\section{Modular Safety Filter}\label{Section: Modular Safety Filter}

\begin{figure}[t]
    \centering
    \includegraphics[trim=16cm 0cm 0cm 0cm, width=1\linewidth]{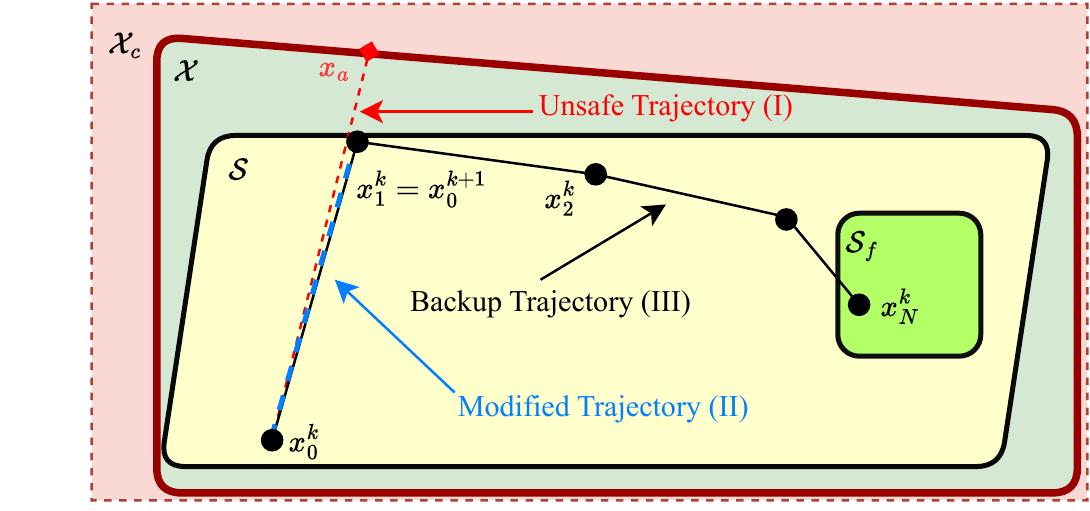}
    \caption{At the time $k$, an unsafe control input, $u_a(k)$, is received by the safety filter. Since applying this input may result in an unsafe trajectory (I) in the next steps, MSF will find a backup trajectory (III) towards the terminal safe control invariant set, $\mathcal{S}_f$, by applying the safe input, $u_s(k)$. Applying this safe input results in the modified trajectory (II).}
    \label{fig: SF_backup}
\end{figure}

To solve Problem \ref{Problem}, we propose an architecture for which a modular safety filter can be implemented as an independent add-on module by filtering the control signal. Upon receiving $x(t)$ and $u_a(t)$, the safety filter provides the nearest safe input, $u_s(t)$, to the command signal $u_a(t)$, while respecting the system constraints (\ref{equ: constraints}) for all $t\geq0$. The safety filter can be described by the following optimization problem,
\begin{equation} \label{equ: safety filter}
    \begin{aligned}
        u_s(t) & = \argmin_{u} \|u_a(t) - u(t)\|^2_2 \\
        \text{s.t.} &\quad u(t) \in \mathcal{U}, \quad x(t) \in \mathcal{X}, \quad \forall t, \geq 0
    \end{aligned}
\end{equation}
where $||.||_2$ is the 2-norm of a vector. To solve this problem, inspired by \cite{wabersich2021predictive}, we propose a predictive-based safety filter to handle safety constraints at all times, including when potentially unsafe inputs are presented. The predictive safety filter approximates the problem (\ref{equ: safety filter}) by searching for a backup input-state trajectory toward a terminal safe control invariant set with finite-time prediction. The predictive safety filter is outlined below,
\begin{subequations} \label{optimization: SF}
    \begin{equation} \label{optimization: MSF}
        u_s(t) = \argmin_{u_{t}^{k}} \|u_{a}(t)-u_{t}^k\|^2_2
    \end{equation}
    \begin{equation}
        \text { s.t. } \forall k \in \mathcal{N} = \{0,1,2,\cdots,N-1\},
    \end{equation}
    \begin{equation} \label{optimization: model}
        x_{t}^{k+1}=f(x_{t}^{k}, u_{t}^{k}),
    \end{equation}    \begin{equation} \label{optimization: const}
        (x_{t}^{k},u_{t}^{k}) \in (\mathcal{X},\mathcal{U}),
    \end{equation}
    \begin{equation} \label{optimization: final}
        x_{t}^{N} \in \mathcal{S}_f,
    \end{equation}
    \begin{equation} \label{optimization: initial}
        x_{t}^{0}=x(t),
    \end{equation}
\end{subequations}
where (\ref{optimization: MSF}) yields the nearest safe action to $u_a(t)$, (\ref{optimization: model}) is the prediction model, (\ref{optimization: const}) is the admissible set, (\ref{optimization: final}) is the terminal safe control invariant set , (\ref{optimization: initial}) is the initial condition at time $t$, $u_{t}^k$ is the $k^{th}$ element of prediction at time $t$, and $N$ is the prediction horizon. Note that if $u_a(t)$ respects constraints \eqref{optimization: model}-\eqref{optimization: initial}, the safety filter does not alter the input. The solution to this optimization problem yields an input-state backup trajectory at time step $t$ as $(u_{t}^{k},x_{t}^{k})$ for $k \in \mathcal{N}$. The safe set $\mathcal{S}_f$ is a control invariant set and defined to guarantee the recursive feasibility of the filter similar to the terminal condition in MPC \cite{rawlings2017model}, e.g., the equilibrium of point of the system \eqref{equ: nonlinear}. A visual illustration of the state constraints, safe set, terminal safe control invariant set, and the backup trajectory at time $t$ is shown in Fig. \ref{fig: SF_backup}. The safety filter algorithm is summarized in the Algorithm \ref{My_algorithm}.

\begin{Assumption} (Initial Feasibility) \label{Assumption: Initial Feasibility}
    The optimization problem (\ref{optimization: SF}) is feasible at $k=0$.
\end{Assumption}

\begin{Assumption}[Terminal Safe Control Invariant Set , $\mathcal{S}_f$] \label{Assumption: terminal safe control invariant set } The terminal safe control invariant set  $\mathcal{S}_f$ is a known safe control invariant set satisfying $\mathcal{S}_f \subseteq \mathcal{S}$ under the terminal control policy $u_{\mathcal{S}_f}= K_{\mathcal{S}_f}(x)$. \end{Assumption}

Note that the safe set, $\mathcal{S}$, is implicitly considered via the MSF optimization problem (\ref{optimization: SF}). The size of this set depends on the prediction horizon and the size of the terminal safe control invariant set, $\mathcal{S}_f$. Generally, there is no need for a long prediction horizon to achieve a non-conservative solution when only safety is concerned, in contrast to an MPC solution that aims to provide performance and safety. 

\begin{Lemma} [Proof of Safety]
    Let Assumptions \ref{Assumption: Initial Feasibility}-\ref{Assumption: terminal safe control invariant set } hold. Then, the system (\ref{equ: nonlinear}) is safe in the sense of Definition \ref{Def: Safety}.
    %is safe for an infinite time.
\end{Lemma}
\textbf{Proof.} To ensure safety, it is sufficient to prove that the optimization \eqref{optimization: SF} enjoys recursive feasibility. If the optimization \eqref{optimization: SF} admits a solution at $t$, then it means that there exists a sequence of control input $\left\{u_t^0, \ldots, u_t^{N-1}\right\}$ that takes the state trajectory safely inside the control invariant region $\mathcal{S}_f$. Consequently, at $t+1$, one admissible although not optimal solution for \eqref{optimization: SF} always exists, and it is given by $\left\{u_t^1, \ldots, u_t^{N-1}, u_{\mathcal{S}_f}.\right\}$. This is sufficient to ensure the recursive feasibility of \eqref{optimization: SF} and, consequently, the existence of a safe backup trajectory provided by \eqref{optimization: SF} regardless of the attacker's actions. \hfill$\square$

\begin{algorithm} [t]
  \caption{Modular Predictive Safety Filter for CPS}
  \label{My_algorithm}
  \begin{algorithmic}[1]
    \State Initialize $\mathcal{S}_f$, $\mathcal{X}$, $\mathcal{U}$, $N$, $x(0)$, $t=0$.
    \While{true} 
      \State Solve problem (\ref{optimization: SF}) for $u_a(t)$.
      \State Apply $u_s(t)$ to system (\ref{equ: nonlinear}).
      \State Measure system's states, $x(t+1)$, send it to the networked controller and update the initial condition.
            \State $t\xrightarrow{}t+1$
    \EndWhile
  \end{algorithmic}
\end{algorithm}

\textit{Remark 1:} {In this paper, we assumed that the underlying problem is deterministic, nominal, and has zero transmission delay. For other settings, when a simplified model, probabilistic model, or additive disturbance is present, an adaptation of the algorithm is required, which may introduce conservatism; see \cite{wabersich2023data} and\cite{hsu2023safety} for a recent overview of safety filter technology.}

\textit{Remark 2:} It is assumed that the tracking controller does not activate the safety filter in the absence of an attack, ensuring that MSF does not interfere with the anomaly detector unless an attack occurs on the communication channel. To relax this assumption, an additional copy of the MSF can be placed alongside the tracking controller. In this configuration, a pre-filtered control signal is transmitted, while the MSF on the plant side remains inactive in the absence of an attack \cite{9123676}.

\section{Modular Predictive Safety Filter for a Multi-Agent Mobile Robot System}\label{Modular Predictive Safety Filter for Multi-Agent Mobile Robot System}

To evaluate the efficiency of the proposed method on a high-order nonlinear system, we adopt the simulation framework employed in \cite{wang2017safety}, which considered 20 mobile robots. For simplicity, we assume that all mobile robots have the same dynamics and parameters. Unlike \cite{wang2017safety}, which employs linear models, we utilize a nonlinear kinematics model for the $i^{th}$ robot, where $i \in \mathcal{I}$, described as follows:
\begin{equation} \label{equ: Kinematics}
    \dot{x}_i=v_i \cos \theta_i, \quad \dot{y}_i=v_i \sin \theta_i, \quad \dot{\theta}_i =\omega_i,
\end{equation}
where $x_i$ and $y_i$ represents the position vector $p_i=[x_i,y_i]^{\top}$, $\theta_i$ is the heading, $v_i$ and $\omega_i$ are the control inputs $u_i=[v_i,\omega_i]^{\top}$, and $\mathcal{I} = \{1,2,...,20\}$ is an index set indicating each agent.  

\begin{figure} [t]
    \centering
    \includegraphics[width=1\linewidth]{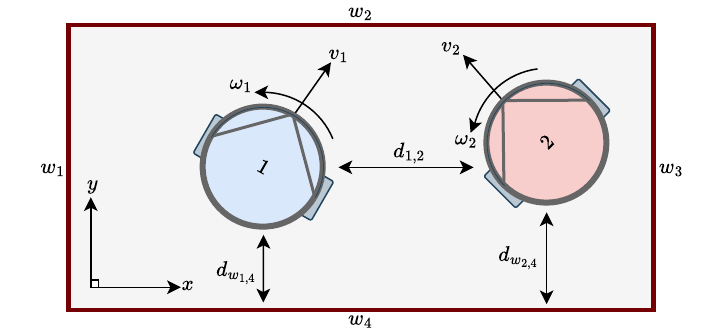}
    \caption{Schematic of mobile robots: linear and angular velocities $(v,\omega)$, and Cartesian coordinates $(x,y)$. The pre-defined safety constraints for the multi-agent system are the distance between two arbitrary robots $d_{i,j}$ and the distance between an arbitrary robot and a wall $d_{w_{i,j}}$.}
    \label{fig: Multi}
\end{figure}

We define the state constraints as the minimum distance between two arbitrary agents ($d_{[i,j]}\geq\delta_a$) and the minimum distance between one arbitrary agent and walls ($d_{w_{[i,j]}}\geq\delta_w$), described by (\ref{equ: admissible set mobile robot}). Graphical visualization of these constraints for two agents, as well as the inertial frame, is depicted in Fig. \ref{fig: Multi}. Note that constraint set (\ref{equ: admissible set mobile robot}) is user-defined and only specifies state constraints. The safe set is also dependent on the velocities and actuator limitations, which are implicitly considered by the MSF.
\begin{subequations} \label{equ: admissible set mobile robot}
    \begin{equation}
        d_{[i,j]} := \{ ||p_i-p_j||_2 : \forall i ,j \in \mathcal{I} , i \neq j \}, \\
    \end{equation}
    \begin{equation}
        d_{w_{[i,j]}} := \{ ||p_i-w_j||_2 : \forall i \in \mathcal{I} , \forall j \in \mathcal{J} \},
    \end{equation}
\end{subequations}
where, $w_j$ is the position of the $j^{th}\in \mathcal{J}:=\{1,2,3,4\}$ wall, $d_{[i,j]}$ represents the distance between $i^{th}$ and $j^{th}$  agents, $d_{w_{[i,j]}}$ represents the distance between $i^{th}$ agent and $j^{th}$ wall. 

The terminal safe control invariant set must still be \emph{explicitly} defined. For a \emph{multi-agent mobile system}, we define it as a set of rest points where each robot has zero velocity, and the distance between any two robots exceeds a positive threshold (\ref{optimization: Multi-Agent const}-\ref{optimization: Multi-Agent velocity}). To avoid the collision, we define constraints over the backup trajectories for each agent as a minimum distance between the backup trajectories over the prediction horizon. The safety filter must be able to find safe backup trajectories that do not collide and have zero velocity at the end of the prediction horizon. We emphasize that this design is not case-dependent for mobile robots; it can be applied to similar scenarios, such as a group of aerial robots or any multi-agent system that has an equilibrium point. A circular reference trajectory with constant radius, $r_0$, and constant angular velocity, $\omega_0$, is defined as the formation task for the multi-agent mobile robot system as follows:
\begin{equation} \label{REF}
    \begin{array}{l}
    x_i^d =  r_0 \sin (w_0 t + \frac{2\pi}{|\mathcal{I}|} (i-1)), \\
    y_i^d =  r_0 \cos (w_0 t + \frac{2\pi}{|\mathcal{I}|} (i-1)), \\
    \end{array}
\end{equation}
where $|\mathcal{I}|$ is the cardinality of $\mathcal{I}$. The modular safety filter for the multi-agent mobile robot system is defined as follows:

\begin{subequations} \label{optimization: Multi-Agent}
    \begin{equation} \label{optimization: Multi-Agent Obj}
        U_s = \argmin_{U_{t}^{k}} ||U_{a}-U_{0}^k||^2_2
    \end{equation}
    \begin{equation} \label{optimization: Prediction horizon}
        \text { s.t. } \forall k \in \mathcal{N} = \{0,1,2,\cdots,N-1\},
    \end{equation}
    \begin{equation} \label{optimization: Multi-Agent model}
        X_{t}^{k+1}=f_d(X_{t}^{k}, U_{t}^{k}),
    \end{equation}
    \begin{equation} \label{optimization: Multi-Agent const}
        U_{t}^{k} \in \mathcal{U},
    \end{equation}
    \begin{equation}  \label{optimization: Multi-Agent Initial}
        X_{t}^{0}=X(t),
    \end{equation}
    \begin{equation}  \label{optimization: Multi-Agent backup}
        {d_{t}^{k}}_{[i,j]} \geq \delta_a, \forall i \in \mathcal{I}, \quad \forall j \in \mathcal{J}, \quad \forall k \in \mathcal{N},
    \end{equation}
    \begin{equation}  \label{optimization: Multi-Agent backup-wall}
        {{d_w}_{t}^{k}}_{[i,j]} \geq \delta_w, \forall i \in \mathcal{I}, \quad \forall j \in \mathcal{J}, \quad \forall k \in \mathcal{N},
    \end{equation}
    \begin{equation}  \label{optimization: Multi-Agent velocity}
        {d_{t}^{N}}_{[i,j]} \geq \delta_a, \quad {{d_w}_{t}^{N}}_{[i,j]} \geq \delta_w, \quad {v}_{t}^{N} = 0,
    \end{equation}    
\end{subequations}
where ${X}$ and ${U}$ are the stacked states and control inputs for all agents defined as $X=[x_1,y_1,\theta_i,..., x_{20},y_{20},\theta_{20}]^{\top}$ and $U=[u_1,v_1, ..., u_{20},v_{20}]^{\top}$, respectively. Additionally, \eqref{optimization: Multi-Agent model} represents the discrete form of \eqref{equ: Kinematics} for the multi-robot system with the time step $T_s^{s}$ as follows:
\begin{equation} \label{eq:F_d_Multi-Agent} f_d(X(t), U(t)) = \begin{bmatrix}
    x_1(t)\\y_1(t)\\\theta_1(t)\\ \vdots \\ x_{20}(t)\\y_{20}(t)\\\theta_{20}(t)
\end{bmatrix} + T_s^{s} \begin{bmatrix} v_1(t) \cos \theta_1(t) \\ v_1(t) \sin \theta_1(t) \\ w_1(t) \\ \vdots \\ v_{20}(t) \cos \theta_{20}(t) \\ v_{20}(t) \sin \theta_{20}(t) \\ w_{20}(t) \end{bmatrix}. \end{equation}
Also, ${d_{t}^{k}}_{[i,j]}$ and ${{d_w}_{t}^{k}}_{[i,j]}$ are the distance similar to (\ref{equ: admissible set mobile robot}) at time $t$ and $k^{th}$ prediction element. Equations (\ref{optimization: Multi-Agent backup}-\ref{optimization: Multi-Agent velocity}) are defined for the backup trajectory to avoid collisions.

\section{Numerical Results}\label{section: Numerical Results}

To evaluate the effectiveness of the proposed method, two attack scenarios are implemented, following the intelligent attack definition in section \ref{Attacker Abilities} and safety filter setup in section \ref{Modular Predictive Safety Filter for Multi-Agent Mobile Robot System}. Note that the settings for the safety filter are identical for both scenarios, as the safety filter does not depend on the type of attack. A description of the simulation setup and safety filter can be found in Table \ref{table: I}. This simulation takes $15$ seconds, and attacks are applied in \( t \in [5, 10] \, \mathrm{sec} \). We also employed an MPC controller for the tracking problem with a predictive horizon equal to $100$, representing function $u_c(t) = g(r(t),x_a(t))$ in section \ref{tracking controller}. Note that the safety filter uses a shorter prediction horizon of $N=3$, which is sufficient for safety. This demonstrates the practicality of using a short-horizon safety filter for local policies, accommodating the plant's computational limits while leveraging a large-horizon tracking controller where resources permit.

\begin{table}[htbp]
  \centering
  \caption{Simulation Parameters}
  \label{table: I}
  \renewcommand{\arraystretch}{1.1} % Increase vertical spacing
  \begin{tabular}{ |c|c|c|c| }
    \multicolumn{4}{l}{\textbf{Modular Predictive Safety Filter}} \\
    \hline \hline
    \textbf{Parameter} & \textbf{Value} & \textbf{Parameter} & \textbf{Value} \\  
    \hline
    \(N\) & 3 & \(T_s^{\text{filter}}\) & $0.02$ [sec] \\ 
    \hline
    \(\delta_a\) & $0.2$ [m] & \( \delta_w\) & $0.2$ [m]\\ 
    \hline        
    \hline
  \end{tabular}
  \quad
  \begin{tabular}{ |c|c|c|c| }
    \multicolumn{4}{l}{\textbf{Multi-Agent Mobile Robot System}} \\
    \hline \hline
    \textbf{Parameter} & \textbf{Value} & \textbf{Parameter} & \textbf{Value} \\  
    \hline
    \(w_0\) & \( 0.4 \, [\frac{ \text{rad}}{ \text{s}}]\) & $|\mathcal{I}|$ & 20 \\ 
    \hline
    \(r_0\) & \(1.5 \, \text{[m]}\) & \(T_s^{s}\) & 0.02 [sec] \\
    \hline
    \(v_{\text{min}}\) & \(-2 \, [\frac{m}{s}]\) & \(v_{\text{max}}\) & \(+2 \, [\frac{m}{s}]\) \\ 
    \hline 
    \(\omega_{{\text{min}}}\) & \(-2 \,[\frac{ \text{rad}}{ \text{s}}]\) & \(\omega_{{\text{max}}}\) & \(+2 \, [\frac{ \text{rad}}{ \text{s}}]\) \\ 
    \hline 
    \(x_{\text{min}}\) & \(-2 \, \text{[m]}\) & \(x_{\text{max}}\) & \(+2 \, \text{[m]}\) \\ 
    \hline
    \(y_{\text{min}}\) & \(-2 \, \text{[m]}\) & \(y_{\text{max}}\) & \(+2 \, \text{[m]}\) \\     
    \hline 
    \hline
  \end{tabular}
\end{table}

 The anomaly detector in the networked controller layer is defined as:  
\begin{equation} \label{eq: anomaly detector-sim}  
    a(t) = \begin{cases}  
        1 & \text{if } \| X_a(t) - X_c(t) \| \geq \varepsilon, \\  
        0 & \text{otherwise},  
    \end{cases}  
\end{equation}  
where \( X_c(t) = f_d(X_a(t-1), U_c(t-1)) \) represents the expected state after applying \( U_c \), and \( \varepsilon = 10^{-6} \) is the detection threshold, which can be set as small as the solver's numerical precision. This means that if the system's response deviates slightly from the expected state, $X_c(t)$, the anomaly detector is triggered. We emphasize that \eqref{eq: anomaly detector-sim} is used solely for simulation purposes, and any other anomaly detector can be employed.

\subsection{First scenario: Intelligent attack}
\begin{figure}[ht]
    \centering
    \includegraphics[width=1\linewidth]{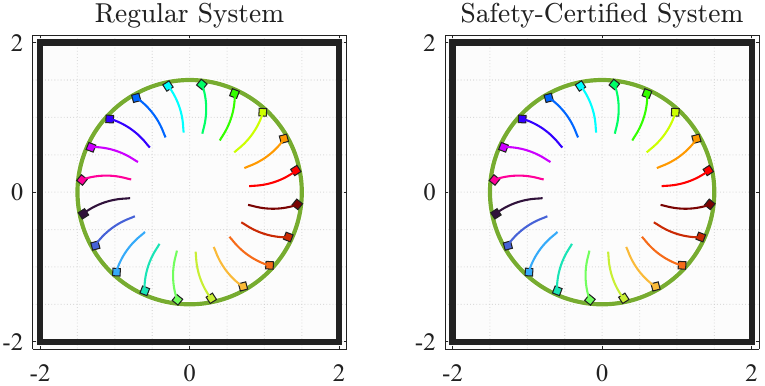}
    \caption{The multi-agent mobile robot system assigned to a formation task: following a circular trajectory at $t= 0.1  \, \mathrm{sec} $. (Before the intelligent attack)}
    \label{fig: Before attack}
\end{figure}

\begin{figure}[ht]
    \centering
    \includegraphics[width=1\linewidth]{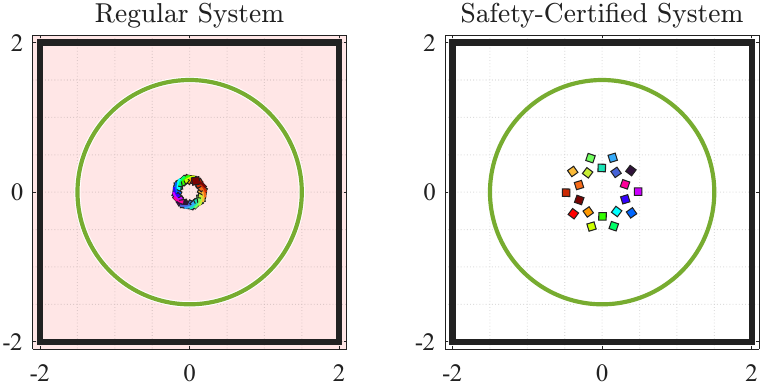}
    \caption{The multi-agent mobile robot system assigned to a formation task: following a circular trajectory at $t= 8  \, \mathrm{sec} $. (During the intelligent attack)}
    \label{fig: During attack}
\end{figure}

\begin{figure}[ht]
    \centering
    \includegraphics[width=1\linewidth]{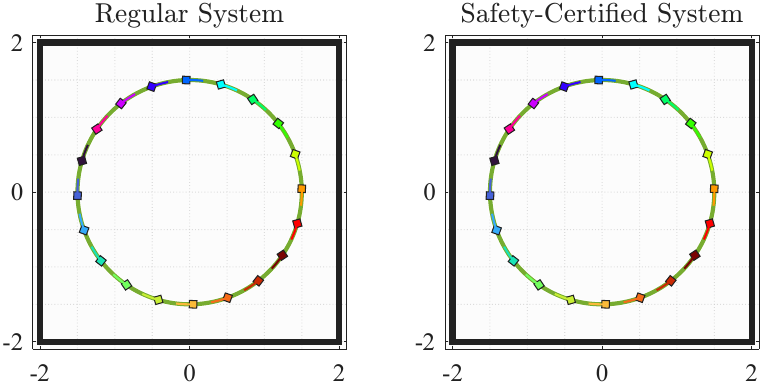}
    \caption{The multi-agent mobile robot system assigned to a formation task: following a circular trajectory at $t= 15  \, \mathrm{sec} $. (After the intelligent attack)}
    \label{fig: After attack}
\end{figure}

Let the attacker read and manipulate the control and sensor measurement signals to perform an undetectable covert attack \cite{smith2015covert}, for $t \in [5,10] \, \mathrm{sec} $. The attack vector on the measurement signals is described as follows:
\begin{equation} \label{equ: Intelligent Attack-state}
    X_a(t+1) =  f_d(X_a(t),U_c(t)),
\end{equation}
where $X_a(t)$ is equal to the system's state $X(t)$ at $t=5 \, \mathrm{sec}$. The evolution of equation (\ref{equ: Intelligent Attack-state}) provides a state trajectory that the anomaly detector expects to see in the networked controller based on the control signal, $U_c$. On the other hand, the attack control input, $U_a$, is computed via an optimization problem whose objective is to cause a collision at the origin. The results of simulation using the CasADi toolbox \cite{andersson2019casadi} for three snapshots at $t= 0.1  \, \mathrm{sec} $, $t= 8  \, \mathrm{sec} $, $t= 15  \, \mathrm{sec} $ representing before, during, and after the attack period are shown in Fig.(\ref{fig: Before attack}-\ref{fig: After attack}). The regular system, shown on the left, has no safety mechanism, while the safety-certified system, shown on the right, uses the filter defined via Algorithm (\ref{My_algorithm}). A video of this simulation is presented \href{https://www.youtube.com/watch?v=kBO05D3sZiE}{here}\footnote{\href{https://www.youtube.com/watch?v=kBO05D3sZiE}{Intelligent Attack: https://www.youtube.com/watch?v=kBO05D3sZiE}}.

As depicted in Fig. \ref{fig: Before attack}, the safety filter has no impact on the formation task before the attack, and agents converge to the circular trajectory (\ref{REF}), shown by the green circle, from their initial conditions. Fig. \ref{fig: During attack} illustrates the impact of the safety filter during the attack, which maintains the system's safety and prevents collisions. Finally, Fig. \ref{fig: After attack} demonstrates how the system can recover itself once the attack is finished.


\begin{figure}[ht]
    \centering
    \includegraphics[width=1\linewidth]{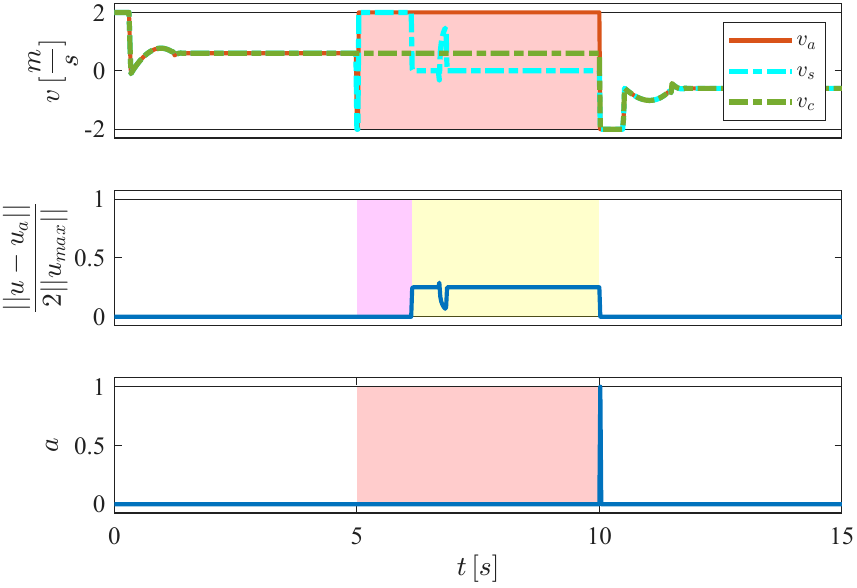}
    \caption{Effect of the proposed modular safety filter on the first agent under an intelligent attack. Here, \(v\) represents the translational velocity, and \(\frac{||u - u_s||}{2 ||u_{\text{max}}||}\) denotes the normalized control input vector, and $a$ is the value of the anomaly detector. (Intelligent attack)}
    \label{fig: Anomaly_collision}
\end{figure}

The first subplot in Fig. \ref{fig: Anomaly_collision}, denoted as $v$, displays the safety-certified and attack inputs for the first agent. For $t \in [0,6.14] \,  \cup [10,15] \, \mathrm{sec} $, the first agent remains safe since there is no modification to the input. It is important to note that for $t \in [5,6.14] \, \mathrm{sec} $, demonstrated by the magenta area in the second subplot, attack input is applied; however, there is no modification as the corresponding agent remains safe. For $t \in [6.14,10] \, \mathrm{sec} $, as depicted by the yellow area, there is a significant correction by the filter to prevent unsafe situations. During $t \in [10,15] \, \mathrm{sec} $, after the attack period, the system successfully recovers to its normal condition, and the safety filter has zero impact. It should be noted that the anomaly detector, \eqref{eq: anomaly detector-sim}, cannot detect the intelligent attack for \( t \in [5, 10] \, \mathrm{sec} \), and it activates only after the attack has finished, see the third subplot in Fig. \ref{fig: Anomaly_collision}. This is due to the nature of the intelligent attack, which exploits system dynamics and knowledge of control input signal $U_c$ to generate data that the anomaly detector expects to observe, i.e., $X_c(t)=X_a(t)$, resulting in an undetectable attack.

\begin{figure}[ht]
    \centering
    \includegraphics[width=1\linewidth]{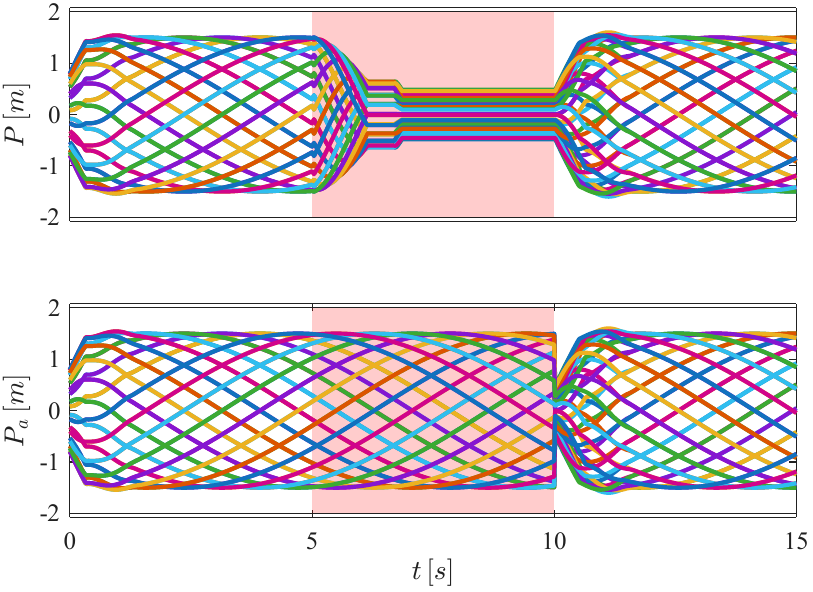}
    \caption{Position of agents through time: \( P \) represents the actual trajectory of the plant, while \( P_a \) denotes the trajectory received by the networked controller under attack-free and intelligent attack conditions. (Intelligent attack)}
    \label{Trajectory_collision}
\end{figure}

The actual position of the system \eqref{equ: Kinematics}, denoted by \( P(t) =[x_1,y_1, \hdots, x_{20},y_{20}]^\top\), and the potentially attacked position received by the networked controller \( P_a(t) \) are illustrated in Fig \ref{Trajectory_collision}. For \( t \in [0,5] \, \mathrm{sec} \), no attack occurs, and the system operates normally, i.e., \( P(t) = P_a(t)\). During \( t \in [5,10] \, \mathrm{sec} \), the attacker attempts to drive all robots to the origin, causing a collision (as shown in the upper subplot), while generating states using \eqref{equ: Intelligent Attack-state} to simulate normal conditions and evade detection. At \( t = 10 \, \mathrm{sec} \), when the attack ends, the anomaly detector observes a sudden jump in the plant's states and identifies the attack; see the bottom subplot in Fig. \ref{Trajectory_collision} and \ref{fig: Anomaly_collision}. Despite this, the safety filter successfully prevents the collision by stopping the robots, as evidenced by \( P(t) \) in Fig. \ref{Trajectory_collision}.


\subsection{False Data Injection Attack}

\begin{figure}[ht]
    \centering
    \includegraphics[width=1\linewidth]{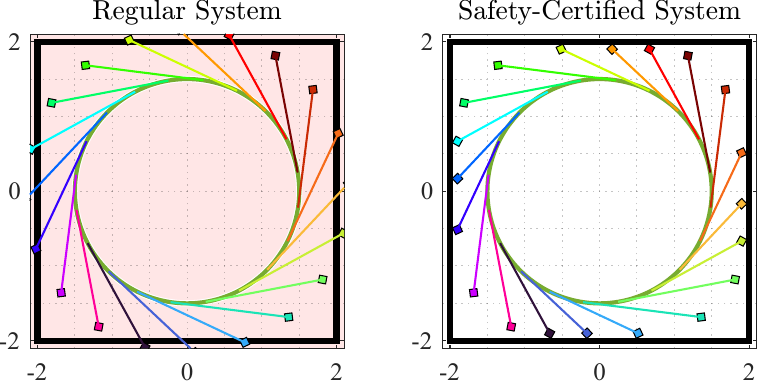}
    \caption{The multi-agent mobile robot system assigned to a formation task: following a circular trajectory at $t= 5.49  \, \mathrm{sec} $. (During the FDI attack)}
    \label{fig: During attack FDI}
\end{figure}

For all agents, we consider an FDI attack where $u_a (t) = -u_c(t)+[v_{max},0]^\top$ for all agents. The objective of this attack is to steer all the agents outside of the admissible set, $\mathcal{X}$, as shown by the black square. The results are shown in Fig. \ref{fig: During attack FDI} at $t=5.49  \, \mathrm{sec} $. The safety-certified system prevents the agents from leaving the admissible set by forcing them to stop before reaching the boundaries. Since the remaining results are similar to the intelligent attack, they are not included in this paper. For further details, please visit the \href{https://www.youtube.com/watch?v=cprja-LznkI}{Link}\footnote{\href{https://www.youtube.com/watch?v=cprja-LznkI}{FDI Attack: https://www.youtube.com/watch?v=cprja-LznkI}}

\section{Conclusions and Future Works}\label{Discussion and Concluding Remarks}

This paper presents a modular safety filter for cyber-physical systems designed to ensure safety at all times, including in the presence of attacks. The safety filter ensures safety regardless of whether the received control command is safe or compromised by an attack, making it effective across various attacks without any assumptions on the attack model. Demonstrating the separation of safety and performance criteria, the proposed solution allows for safety during attacks alongside any high-performance controller. This highlights the versatility of safety filters in cyber-physical system applications, especially given that constrained controllers like MPC cannot optimize all types of cost functions.

The proposed safety filter is inspired by predictive safety filters developed for learning control. This paper illustrates the effectiveness of a modular approach to the safety of CPS that can handle nonlinear and high-order systems. Depending on the system's characteristics, alternative safety filter solutions proposed for learning control can likely be used with minor adjustments to CPS. These methods include control barrier functions and Hamilton-Jacobi analysis, where their extensions can account for uncertain, time-delay, and stochastic settings \cite{wabersich2018linear,ames2019control,bansal2017hamilton}. Each method comes with its advantages and disadvantages. Still, it is worth noting that calculating safe sets and backup trajectories is not as straightforward in other methods as in predictive filters, where they are calculated implicitly with the cost of solving an on-the-fly optimization.

We emphasize that the proposed safety filter can be adapted for a distributed scenario if each agent can communicate with its neighbors or predict their behavior. Developing a distributed version of the proposed method is our next step to enhance practicality and reduce computational complexity \cite{muntwiler2020distributed}. A critical consideration when using safety filters in a non-deterministic setting is their tendency to introduce conservatism. If an accurate model of the system is unavailable, an extremely short prediction horizon and a small final set are adopted, or if there is a significant delay, safety filters may introduce unnecessary caution as any robust, constrained solution. However, since the proposed modular solution is implemented as an add-on that is unaware of the tracking controller and attacks, any conservatism in the safety filter may cause the anomaly detector to detect an attack incorrectly. Further work is required to establish how conservatism affects anomaly detectors in the networked controller, how the impact of conservatism on the anomaly detector can be mitigated, and whether communication between the modules may be required.

\bibliographystyle{IEEEtran}
\bibliography{references}
\balance
\end{document}